\documentclass[preprint,aps]{revtex4}
\usepackage{setspace}
\usepackage{graphicx}
\begin{document}
\begin{center}
{\bf \large Broadband Dielectric Spectroscopy of Ruddlesden-Popper Sr$_{n+1}$Ti$_{n}$O$_{3n+1}$ ($n$ = 1, 2, 3) Thin Films}
\end{center}
\begin{singlespace}
\begin{tabbing}
\hspace{.25in} \=
{\small N. D. Orloff$^{1, 2}$, W. Tian$^3$, C. J. Fennie$^4$, C.-H. Lee$^{3,5}$, D. Gu$^2$, J. Mateu$^6$, X. X. Xi$^5$,}\\
\> K. M. Rabe$^7$, D. G. Schlom$^3$, I. Takeuchi$^{1}$, and J. C. Booth$^2$\\ \\
\>{\scriptsize \it $^1$Department of Materials Science and Engineering and Department of Physics, University of Maryland,}\\
\>{\scriptsize \it ~College Park, Maryland 20742, USA}\\
\>{\scriptsize \it $^2$National Institute of Standards and Technology, Boulder, Colorado 80305, USA}\\
\>{\scriptsize \it $^3$Department of Materials Science and Engineering, Cornell University, Ithaca, New York 14853-1501, USA}\\
\>{\scriptsize \it $^4$School of Applied and Engineering Physics, Cornell University, Ithaca, New York 14853-1501, USA}\\
\>{\scriptsize \it $^5$Department of Materials Science and Engineering, The Pennsylvania State University, University Park,}\\
\>{\scriptsize \it ~Pennsylvania 16802-5005}\\
\>{\scriptsize \it $^6$Department of Signal Theory and Communications, Universitat Polit$\grave{e}$cnica de Catalunya, 08034 Barcelona, Spain}\\
\>{\scriptsize \it $^7$Department of Physics and Astronomy, Rutgers University, Piscataway, New Jersey 08854-8019, USA}\\
\end{tabbing}
\end{singlespace}
\begin{center}
\begin{bf}
Abstract
\end{bf}
\end{center}

\small
We explore the frequency-dependent relative permittivity of Ruddlesden-Popper series Sr$_{n+1}$Ti$_{n}$O$_{3n+1}$ ($n$ =~1,~2,~3) thin films as a function of temperature and dc electric field. Interdigitated capacitors and coplanar waveguides were used to extract the frequency response from 500 Hz to 40 GHz. At room temperature, the in-plane relative permittivities ($\epsilon_{11}$) obtained for Sr$_{n+1}$Ti$_{n}$O$_{3n+1}$ ($n$ =~1,~2,~3) were 42$\pm$3, 54$\pm$3, and 77$\pm$2, respectively, and were independent of frequency. At low temperatures, $\epsilon_{11}$ increases and electric field tunability develops in Sr$_{4}$Ti$_{3}$O$_{10}$.
\newpage
The Ruddlesden-Popper (RP) homologous series Sr$_{n+1}$Ti$_{n}$O$_{3n+1}$ is a unique set of materials whose dielectric properties have yet to be fully explored and have the potential for far reaching applications much like SrTiO$_3$, the $n$ = $\infty$ member of this series \cite{RUDDLESDEN:1957zv}. Series members $n$ = 1, 2, 3 are predicted to have fairly high relative permittivities \cite{FennieS2TO4}. Although historically RP materials have been the focus of a variety of theoretical \cite{FennieS2TO4,MusicRPSTO,NogueraRPSTO} and experimental \cite{WisePerovskites,Wise:2001fv,Kamba:2003zl,SohnRPSTO} research, until the recent growth of single-phase RP thin-films \cite{Madhavan:1996vn,Konishi:1998hl,Jia:1999xe,Iwazaki:1999gd,RPSchlom,RPSchlom2,Burriel:2007so,Shibuya:2008zp,Okude:2008zp}, detailed dielectric measurements of single-phase materials have been challenging \cite{Noujni:2004qc}. The synthesis of stable, single-phase RP thin films opens new possiblities for applications in multilayer heterostructures and suggests experiments to explore the role of their unique crystal structures in ferroelectricity \cite{Fennie:2005zh}.

The Sr$_{n+1}$Ti$_{n}$O$_{3n+1}$ RP series is of particular interest, because SrTiO$_{3}$ is already commonly used in a number of applications: a substrate for perovskite films, integrated nonlinear compensation devices \cite{STOJordi}, and to enhance ferroelectricity \cite{PbSTO}. When strained, SrTiO$_{3}$ can exhibit ferroelectricity even at room temperature and at modest bias fields \cite{StrainSTOSchlom}. This increase in tunability unfortunately signifies a remarkable increase in both the relative permittivity and loss, rendering them less useful for high-frequency applications. The Sr$_{n+1}$Ti$_{n}$O$_{3n+1}$ RP series may provide a new class of tunable materials which have decreased loss, yet still retain many of the characteristics that make SrTiO$_{3}$ so useful.

Typical radio frequency dielectric measurements focus on a single frequency \cite{RPSchlom} or a small range of frequencies, even though frequency dependence carries important information about disorder and relaxation mechanisms. Resonant techniques, such as evanescent microwave microscopy, yield very accurate measurements of local material properties, but are limited in their ability to extract broadband frequency information \cite{LowKEps, Gao:2005kk}. Consequently, lumped element capacitors are often used, an example of which are interdigitated capacitors (IDCs). Distributed effects, which are often overlooked, can compromise results from such devices at high frequencies, where the guided wavelength is comparable to the length of the device \cite{BookChapter}.

Here, we report quantitative measurements of in-plane relative permittivity ($\epsilon_{11}$) for Sr$_{n+1}$Ti$_{n}$O$_{3n+1}$ thin films ($n$ = 1, 2, 3) by means of an ultra-wideband approach \cite{IMSMultilayerOrloff}. To obtain the high-frequency behavior of the RP thin-films and their companion substrate (001) (LaAlO$_3$)$_{0.3}$-(SrAl$_{0.5}$Ta$_{0.5}$O$_3$)$_{0.7}$ (LSAT), we exploited coplanar waveguides (CPWs) rather than IDCs, which were used at frequencies below 200 MHz. By deliberately using distributed devices at high frequencies, we have effectively avoided any errors arising from IDCs. Our single-phase (001) Sr$_{n+1}$Ti$_{n}$O$_{3n+1}$ thin-films were grown by molecular-beam epitaxy (MBE) on 0.5 mm thick (001) LSAT substrates with dimensions 10 mm$\times$10 mm and identical growth conditions to those of Ref. \cite{RPSchlom}. Four-Circle x-ray diffraction measurements revealed that the c-axis lattice constants of the 160 nm - 170 nm thick $n$ = 1, 2, and 3 films were 12.57$\pm$0.02 \AA, 20.42$\pm$0.02 \AA, and 28.05$\pm$0.02 \AA, respectively. The films were commensurate within the experimental error ($\pm$0.0008 \AA). The combined CPW and IDC measurements span 500 Hz to 40 GHz, almost eight decades in frequency.

We fabricated CPWs and IDCs onto the RP thin films and a bare (001) LSAT, substrate using standard lithographic techniques. The electrodes are made from Au approximately 330 nm thick with a 20 nm titanium adhesion layer. The CPWs have prescribed lengths, $\ell$ = \{0.420 mm, 2.155 mm, 3.220 mm, 5.933 mm\}, which improve the measurement accuracy above 200 MHz \cite{MultilineMarks}. Our IDCs also have different active lengths $\ell$ = \{0, 0.100 mm, 1.835 mm, 2.900 mm\}, so they may be fit as a function of length.

The measurements are corrected using on-wafer planar standards patterned on a LaAlO$_3$($\epsilon_{11}\approx\epsilon_{33}$ = 24.6$\pm$0.6 \cite{Harrington:1997rf,Schwab:1997mj}) substrate with identical cross sectional geometries to the test devices. For frequencies below 1 MHz, we measured the complex impedance of the IDCs using an impedance analyzer. Between 1 MHz and 200 MHz, we measured the complex scattering parameters (S-parameters) of the IDCs and from 200 MHz to 1 GHz using a radio frequency vector network analyzer (VNA). At high frequencies, we measured S-parameters of CPWs with a microwave VNA. We extracted the distributed circuit parameters of the bare substrate and thin films for both IDCs and CPWs. After taking the difference between the distributed capacitance of the thin film and substrate, we used finite element simulations of the cross-sectional geometries of our devices to relate the capacitance difference to the respective relative permittivity of the thin film.
\begin{center}
[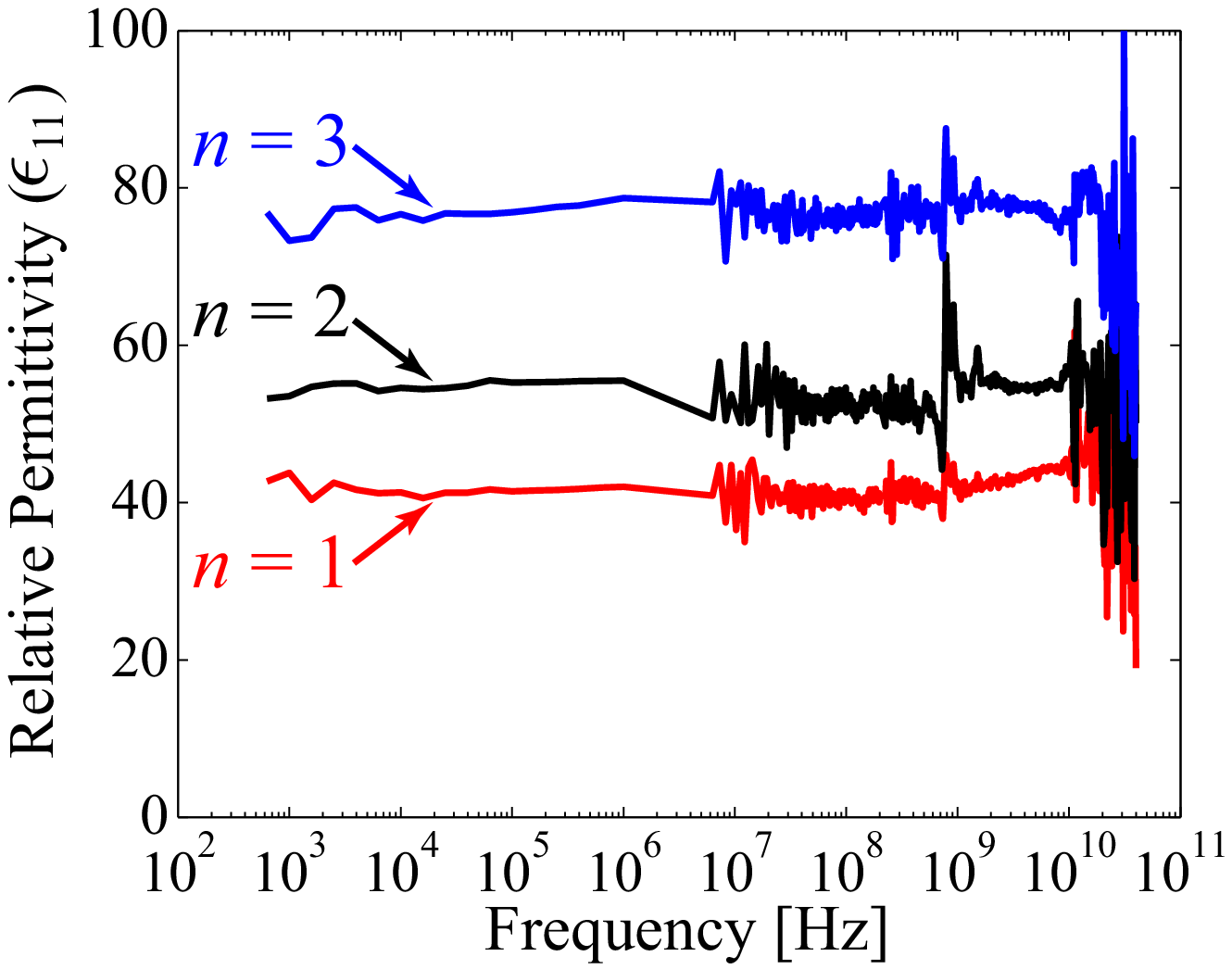]
\end{center}

Figure 1 shows $\epsilon_{11}$ for Sr$_{n+1}$Ti$_{n}$O$_{3n+1}$ ($n$ = 1, 2, 3) thin films and explicitly demonstrates that it is independent of frequency up to 40 GHz. Additionally, we measured the relative permittivity of LSAT ($\epsilon_{11}$ = 23.1$\pm$0.3) over the same measurement frequency range and found it to be consistent with previous reports \cite{TidrowLSAT}. We also found that the present RP materials had relatively low loss tangents, significantly less than 0.04 (the sensitivity of our measurements).
\begin{center}
[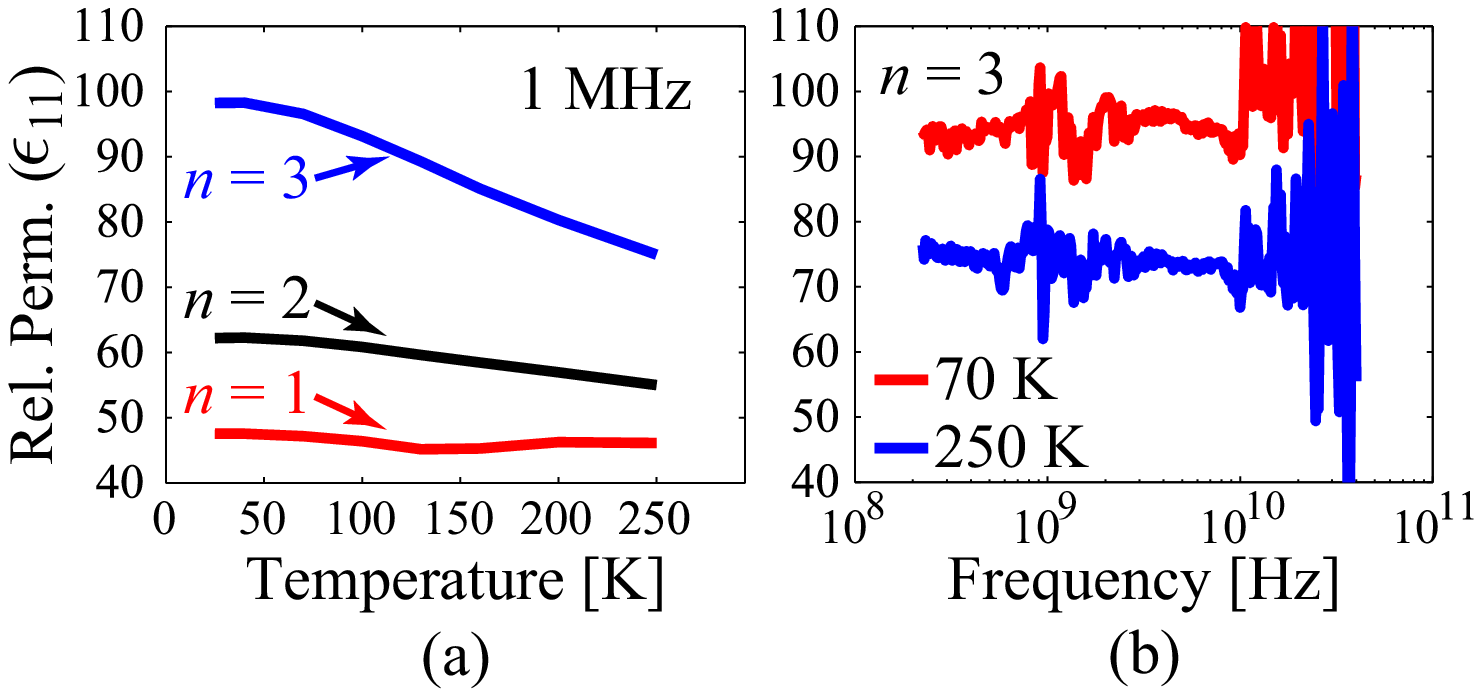]
\end{center}

To further explore the dielectric properties of these materials as a function of temperature, we employed a cryogenic probe station. In Fig. 2(a), we plot the temperature dependence of $\epsilon_{11}$ at 1 MHz for these series members, which we find are consistent with previous reports on the average permittivity in bulk ceramics \cite{Noujni:2004qc}. The $\epsilon_{11}$ for $n$ = 1, 2, 3 increases with decreasing temperature, and this dependence increases with $n$. The temperature dependence seen in Fig. 2(a) for $n$ = 2 and 3 is similar to that of SrTiO$_{3}$ \cite{HembergerSTOTdep}, which is likely due to the temperature dependence of the soft mode \cite{SirenkoPhononsSTO}. In comparison, $n$ = 1 shows relatively weak temperature dependence, a feature that makes Sr$_2$TiO$_4$ attractive as an alternative high-speed, low-loss gate dielectric. Figure 2(b) shows that even at 70 K the high-frequency dependence of Sr$_4$Ti$_3$O$_{10}$ ($n$ = 3) remains flat up to 40 GHz and shows no evidence of any relaxational processes, which could be 
connected with defects or high lattice anharmonics.
\begin{center}
[Table I]
\end{center}

Table I shows the ${\epsilon_{11}}$ averaged over the measurement frequency range for temperatures 295 K, 70 K, and 25 K. Measured results are shown with the values calculated using the first principles technique discussed below. The measured values for 295 K are averaged over the frequency range of 500 Hz - 20 GHz. For 70 K, the mean value is taken from 200 MHz to 20 GHz, and the 30 K result is averaged from 1 kHz to 1 MHz. Experiments on bulk ceramics at room temperature measured $\epsilon_{average}$ = \{37.4, 57.9, 76.1\} for n = 1, 2, 3 \cite{Wise:2001fv,Kamba:2003zl}.

First-principles density functional calculations using projector augmented wave potentials were performed within the local density approximation as implemented in \textsc{vasp} \cite{VASP,PAW}. The wavefunctions were expanded in plane waves up to a kinetic energy cutoff of 500\,eV. Integrals over the Brillouiun zone were approximated by sums on a 6$\times$6$\times$6 $\Gamma$-centered $k$-point mesh. Phonon frequencies were calculated using the direct method where each symmetry adapted mode \cite{bilbao,isotropy} was moved by approximately 0.01\,\AA. Born effective charge tensors were calculated by finite differences of the polarization using the modern theory of polarization \cite{king-smith.prb.93} as implemented in \textsc{vasp}. All structures were fully relaxed.

Interestingly, the three RP systems were found to have highly anisotropic permittivity tensors; for $n$ = (1, 2, 3), $\epsilon_{11}$ = \{62, 100, 150\} and $\epsilon_{33}$ = \{18, 25, 34\}, a result previously overlooked. Although the magnitude of $\epsilon_{11}$ is highly sensitive to constant biaxial strain for reasons previously discussed~\cite{FennieS2TO4}, the relative change as a function of $n$ is consistent with the presented measurements.

The distinct temperature dependence for Sr$_{4}$Ti$_{3}$O$_{10}$ ($n$ = 3) in Fig. 2(b), although not by itself a proof of soft mode temperature dependence, is often an indication of tunability. We, therefore, performed two independent tests using both IDCs and CPWs to measure the dc electric field dependence for this sample. Electric field bias was applied between the center conductor and outer conductor of a 5.933 mm CPW, using a bias tee. Corresponding biased measurements on two IDCs, $\ell$ = \{0.1 mm, 2.9 mm\}, were accomplished using low frequency bias tees, where the bias voltage was applied between the interdigitated fingers of the IDCs.
\begin{center}
[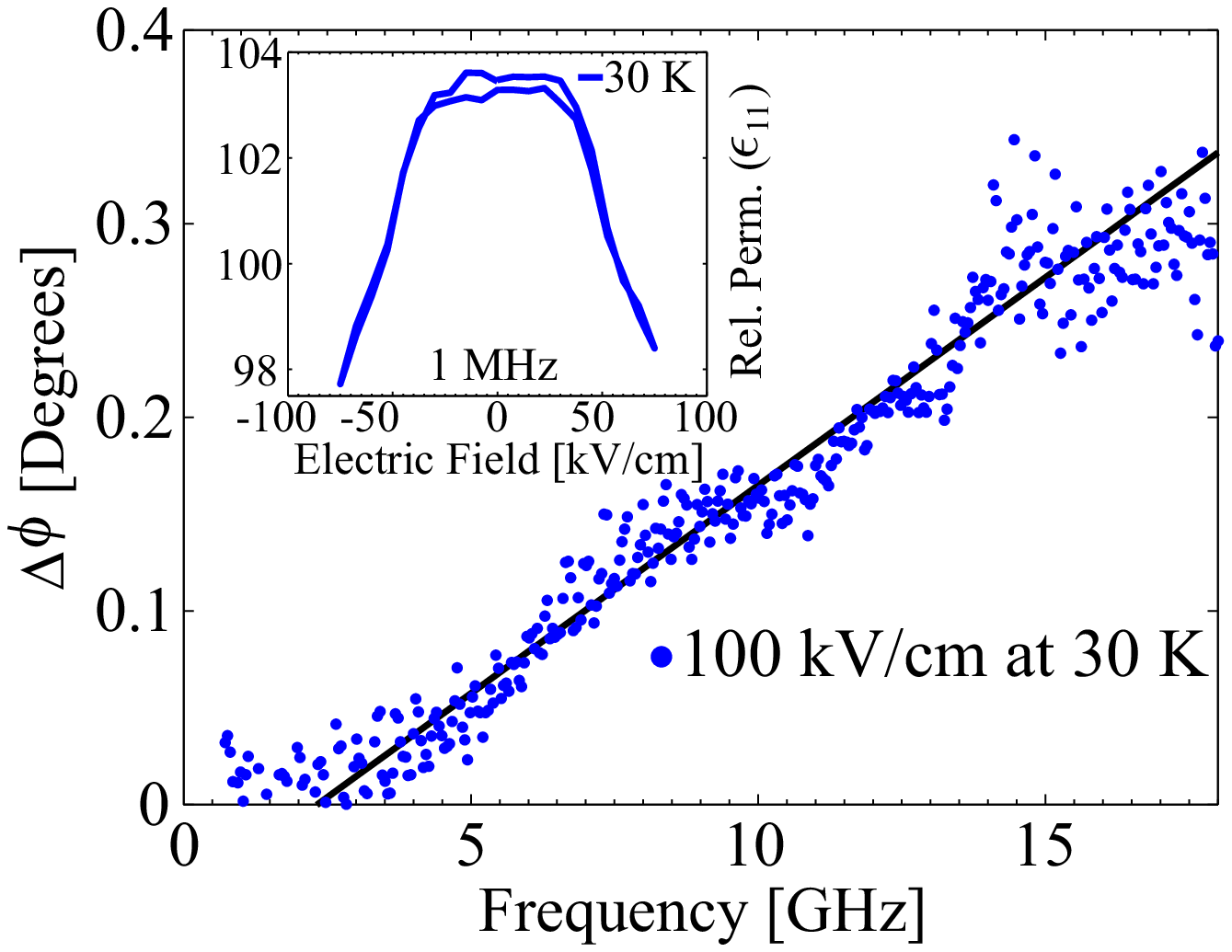]
\end{center}

Figure 3 shows the measured relative change in phase $\Delta \phi$ as a function of frequency for a CPW ($\ell$ = 5.933 mm) on the Sr$_{4}$Ti$_{3}$O$_{10}$ ($n$ = 3) thin film. The measured $\Delta \phi$ for this CPW at 30 K and 100 kV/cm corresponds to a change in film relative permittivity of about 6\%, which is consistent with the low-frequency biased measurements. Both IDC and CPW measurements confirm that the Sr$_{4}$Ti$_{3}$O$_{10}$ ($n$ = 3) thin film can be tuned with modest dc electric fields at low temperatures. For $n$ = 1 and 2, the tunability was less than 3\%, which is approximately the uncertainty in the measurement.

In summary, we have characterized the $\epsilon_{11}$ of the first three members of Sr$_{n+1}$Ti$_{n}$O$_{3n+1}$ RP Series using single-phase epitaxial thin films grown by MBE. The measured $\epsilon_{11}$ was consistent with the value obtained from theory. We have systematically explored the dependence of $\epsilon_{11}$ on $n$, temperature, and dc electric field from 500 Hz to 40 GHz by combining data from IDC and CPW devices. We found evidence that suggests that these dependences are intrinsic to the Sr$_{n+1}$Ti$_{n}$O$_{3n+1}$ phases rather than inclusions of SrTiO$_3$.
\begin{acknowledgments}
The authors thank S. Kamba, A. Lewandowski, H. Romero, D. Novotny, C. Long, and D. I. Orloff for their help in performing this work. We gratefully acknowledge the financial support of the National Science Foundation under grants DMR-0507146, DMR-0820404, DMR-0520471 (MRSEC), and ARO W911NF-07-1-0410.
\end{acknowledgments}
\newpage
\begin{center}
\textbf{References}
\end{center}

\newpage
\begin{center}
\textbf{Tables \& Figure Captions}
\end{center}
\begin{table}[!h]
\begin{smallskip}
\renewcommand{\arraystretch}{.7}
\caption{${\epsilon_{11}}$ for Sr$_{n+1}$Ti$_{n}$O$_{3n+1}$ at 295 K, 70 K, and 30 K.}
\renewcommand{\arraystretch}{1.1}
\centering
\begin{tabular*}{3.25 in}[b]{@{\extracolsep{\fill}}r|lll}
\hline
 & \textbf{\textit{n} = 1} & \textbf{\textit{n} = 2} & \textbf{\textit{n} = 3} \\
\hline
295 K & 42.1$\pm$2.5 & 53.8$\pm$2.9 & 77.2$\pm$2.1 \\
70 K & 49.0$\pm$4.5 & 61.2$\pm$6.8 & 95.5$\pm$4.5 \\
30 K & 47.5$\pm$0.1 & 62.1$\pm$0.1 & 98.1$\pm$0.2 \\
Theory & 62 & 100 & 150 \\
\hline
\end{tabular*} 
\end{smallskip}
\end{table}
\newpage
%
%
\begin{figure}[ht]
\centering
\caption{The frequency dependence (500 Hz to 40 GHz) of the in-plane relative permittivity (${\epsilon_{11}}$) of Sr$_{n+1}$Ti$_{n}$O$_{3n+1}$ ($n$ = 1, 2, 3) thin films at room temperature (296 K).}
\end{figure}
\begin{figure}[!h]
\centering
\caption{(a) Temperature dependence of the in-plane relative permittivity (${\epsilon_{11}}$) for Sr$_{n+1}$Ti$_{n}$O$_{3n+1}$ thin-films with $n$ = 1, 2, and 3. The temperature dependence was extracted from two interdigitated capacitors of lengths $\ell$ = \{0.1 mm, 2.9 mm\}. (b) shows the high frequency dependence of Sr$_{4}$Ti$_{3}$O$_{10}$ ($n$ = 3) thin-film measured with coplanar waveguides at 70 K and 250 K.}
\end{figure}
\begin{figure}[!h]
\centering
\caption{The phase change $\Delta \phi$ of a 5.933 mm coplanar transmission line biased at 100 kV/cm relative to the unbiased line, on the Sr$_{4}$Ti$_{3}$O$_{10}$ ($n$ = 3) thin film. The solid line shows the predicted $\Delta \phi$, a 6\% change in relative permittivity, based on the low frequency response in the inset. The inset depicts the tunability of $\epsilon_{11}$ for Sr$_{4}$Ti$_{3}$O$_{10}$ ($n$ = 3) at 1 MHz and 30 K, which was extracted from two interdigitated capacitors of lengths $\ell$ = \{0.1 mm, 2.9 mm\}.}
\end{figure}
\newpage
\begin{center}
\textbf{Tables \& Figure Captions}
\end{center}
\begin{figure}[!h]
\includegraphics[width= \columnwidth]{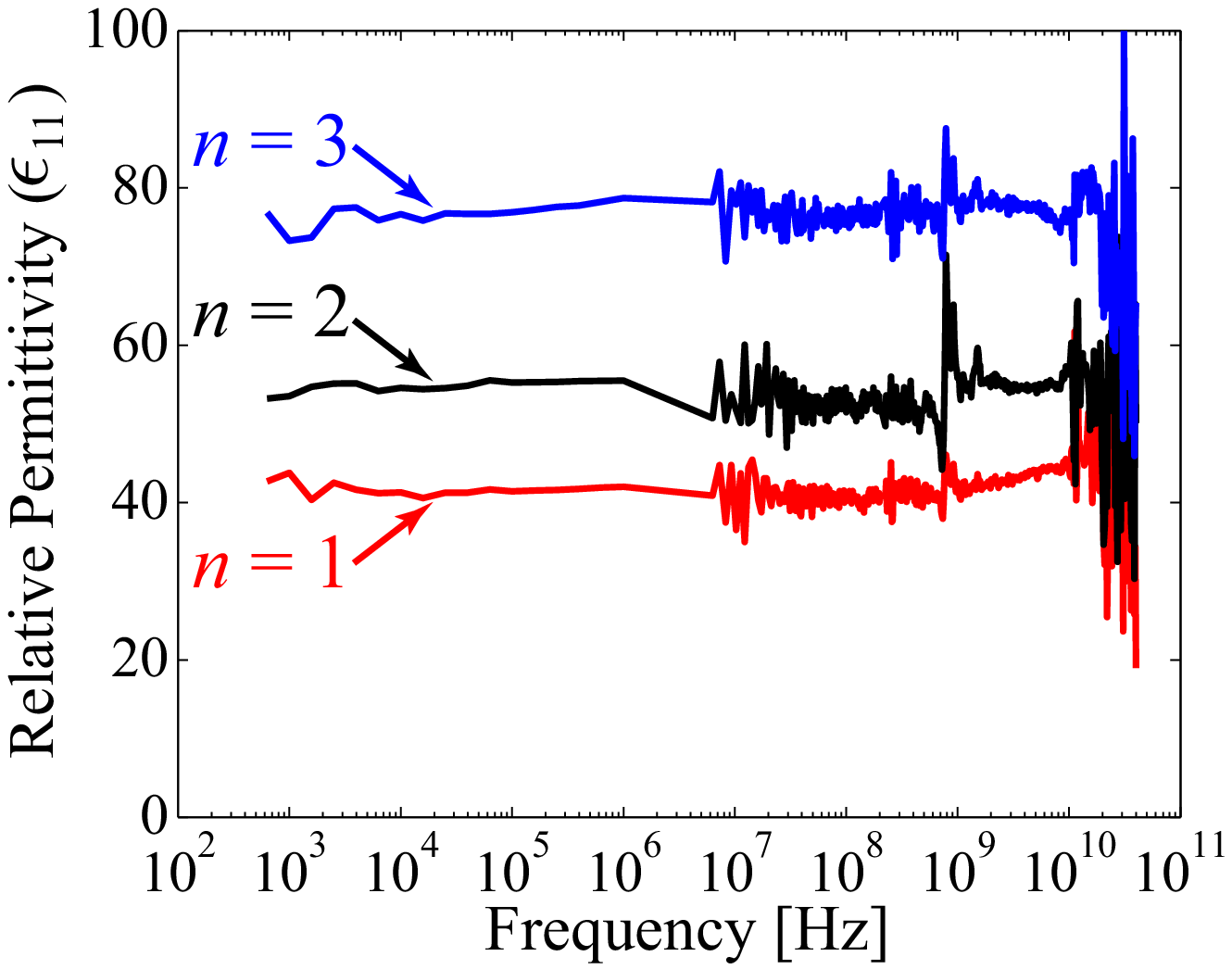}
\end{figure}
\begin{center}
\textbf{Fig. 1}
\end{center}
\newpage
\begin{figure}[!h]
\includegraphics[width= \columnwidth]{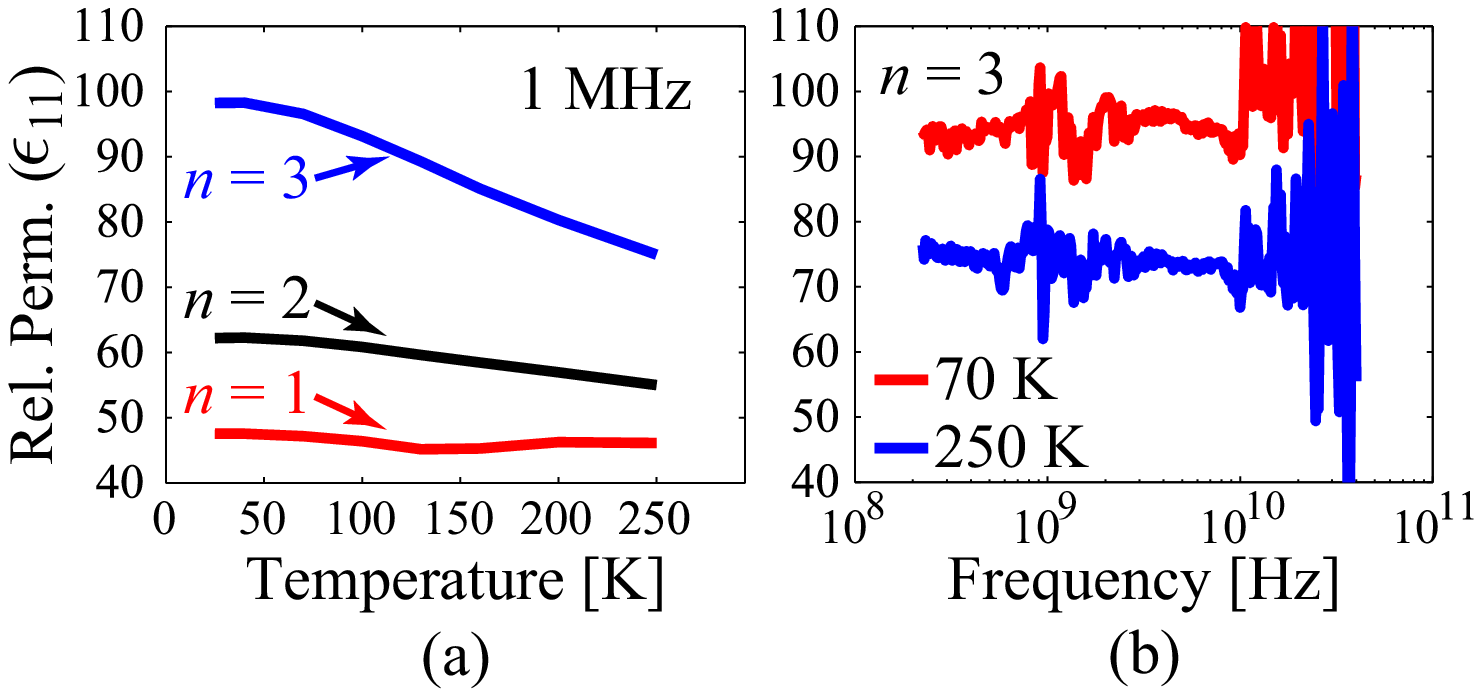}
\end{figure}
\begin{center}
\textbf{Fig. 2}
\end{center}
\newpage
\begin{figure}[!h]
\includegraphics[width= \columnwidth]{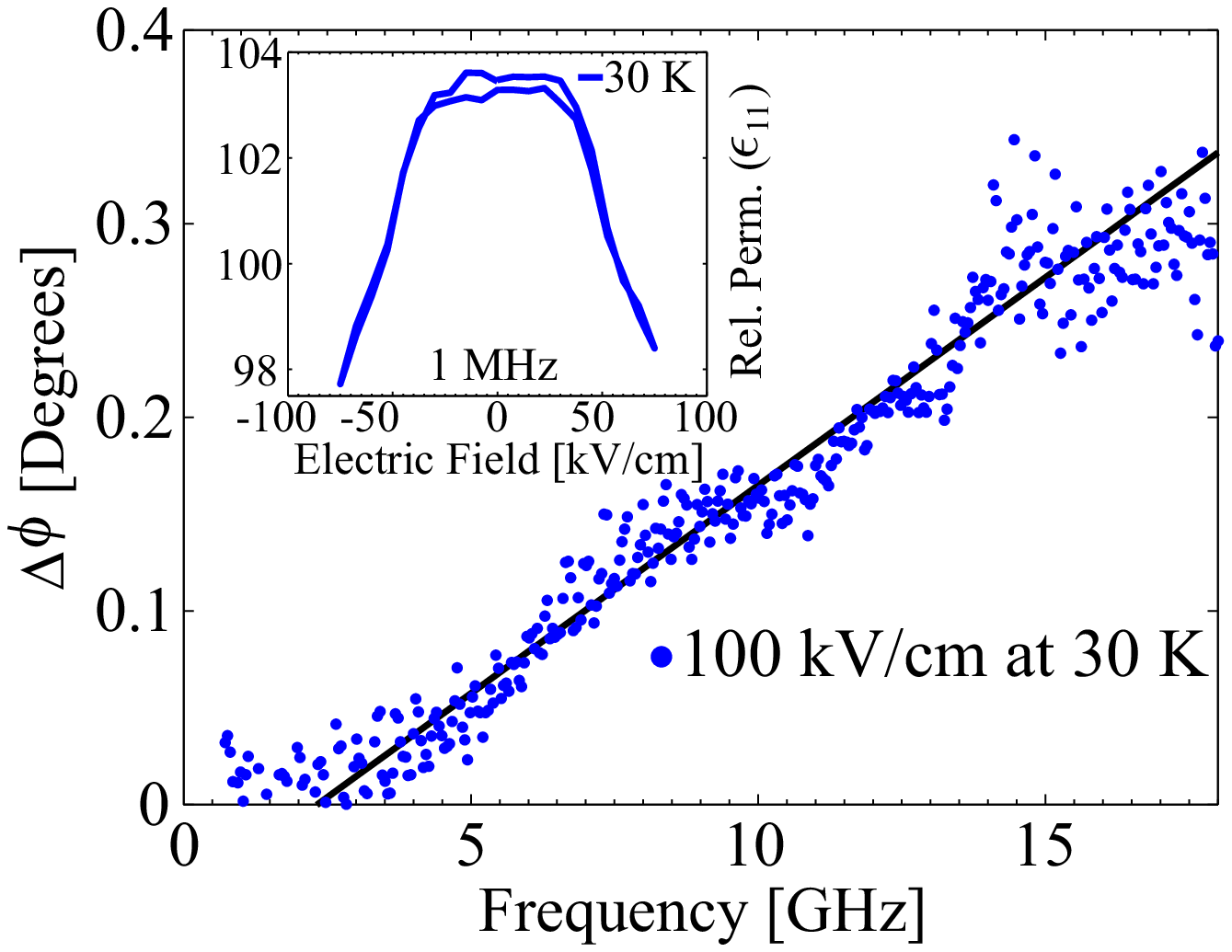}
\end{figure}
\begin{center}
\textbf{Fig. 3}
\end{center}
\end{document}